\documentclass[conference]{IEEEtran}
\IEEEoverridecommandlockouts
\usepackage{tabularx} % 加载 tabularx 宏包
\usepackage{multirow}
\usepackage{cite}
\usepackage{amsmath,amssymb,amsfonts}
\usepackage{algorithmic}
\usepackage{graphicx}
\usepackage{textcomp}
\usepackage{xcolor}
\def\BibTeX{{\rm B\kern-.05em{\sc i\kern-.025em b}\kern-.08em
    T\kern-.1667em\lower.7ex\hbox{E}\kern-.125emX}}
\begin{document}

\title{Wavelet-based  Global-Local Interaction Network with Cross-Attention for Multi-View Diabetic Retinopathy Detection\\
% {\footnotesize \textsuperscript{*}Note: Sub-titles are not captured in Xplore and should not be used}

\author{Yongting Hu$^{a,b}$,Yuxin Lin$^{a,b}$,Chengliang Liu$^{a,b}$,Xiaoling Luo$^{c}$, Xiaoyan Dou$^{d}$, Qihao Xu$^{a,b}$, Yong Xu$^{a,b,*}$ \\
$^{a}$School of Computer Science and Technology, Harbin Institute of Technology, Shenzhen, Shenzhen, China \\
$^{b}$Shenzhen Key Laboratory of Visual Object Detection and Recognition Shenzhen, China \\
$^{c}$College of Computer Science and Software Engineering, Shenzhen University, Shenzhen, China \\
$^{d}$Ophthalmology Department, Shenzhen Second People’s Hospital, Shenzhen, China 
}

%\author{aSchool of Computer Science and Technology, Harbin Institute of Technology, Shenzhen, Shenzhen, China}
\thanks{$*$ Corresponding author: Yong Xu, laterfall@hit.edu.cn}
\thanks{This work was supported in part by the Natural Science Foundation of Guangdong Province under Grant 2025A1515010184 and the project of Guangdong Basic and Applied Basic Research Foundation under Grant 2023B0303000010.}
}

\maketitle

\begin{abstract}
Multi-view diabetic retinopathy (DR) detection has recently emerged as a promising method to address the issue of incomplete lesions faced by single-view DR. However, it is still challenging due to the variable sizes and scattered locations of lesions. Furthermore, existing multi-view DR methods typically merge multiple views without considering the correlations and redundancies of lesion information across them. Therefore, we propose a novel method to overcome the challenges of difficult lesion information learning and inadequate multi-view fusion. Specifically, we introduce a two-branch network to obtain both local lesion features and their global dependencies. The high-frequency component of the wavelet transform is used to exploit lesion edge information, which is then enhanced by global semantic to facilitate difficult lesion learning. Additionally, we present a cross-view fusion module to improve multi-view fusion and reduce redundancy. Experimental results on large public datasets demonstrate the effectiveness of our method. The code is open sourced on https://github.com/HuYongting/WGLIN.
\end{abstract}

\begin{IEEEkeywords}
diabetic retinopathy, multi-view,  global-local interaction, cross-attention
\end{IEEEkeywords}
\vspace{-10pt}
\section{Introduction}

Approximately 537 million adults aged 20-79 in the world suffer from diabetes, and the number is expected to increase to 643 million by 2030 \cite{IDF2021,luo2021mvdrnet}. Diabetic retinopathy (DR) is a common complication of diabetes, and  its advanced stages can lead to irreversible blindness in adults. The severity of DR is primarily determined by the extent of retinal lesions, which are divided into five stages: 0-normal, 1-mild, 2-moderate, 3-severe, and 4-proliferative DR (PDR)\cite{grad2022}. In the advanced stage, the primary lesions include neovascularization  and retinal detachment, which can result in blindness. Therefore, early diagnosis of DR is crucial to prevent vision loss. However, traditional DR detection relies heavily on ophthalmologists, placing significant demands on them, especially as the prevalence of DR continues to rise. Moreover, the subjective nature of the diagnostic process can lead to inconsistent and inaccurate grading of lesions, increasing the risk of misdiagnosis. As a result, automated DR detection is essential.

\begin{figure}[h]  % figure 环境用于插入图片 htbphere, top, bottom, page
	%\captionsetup{belowskip=5pt}
	\centering  % 图片居中显示
	\includegraphics[width=0.5\textwidth]{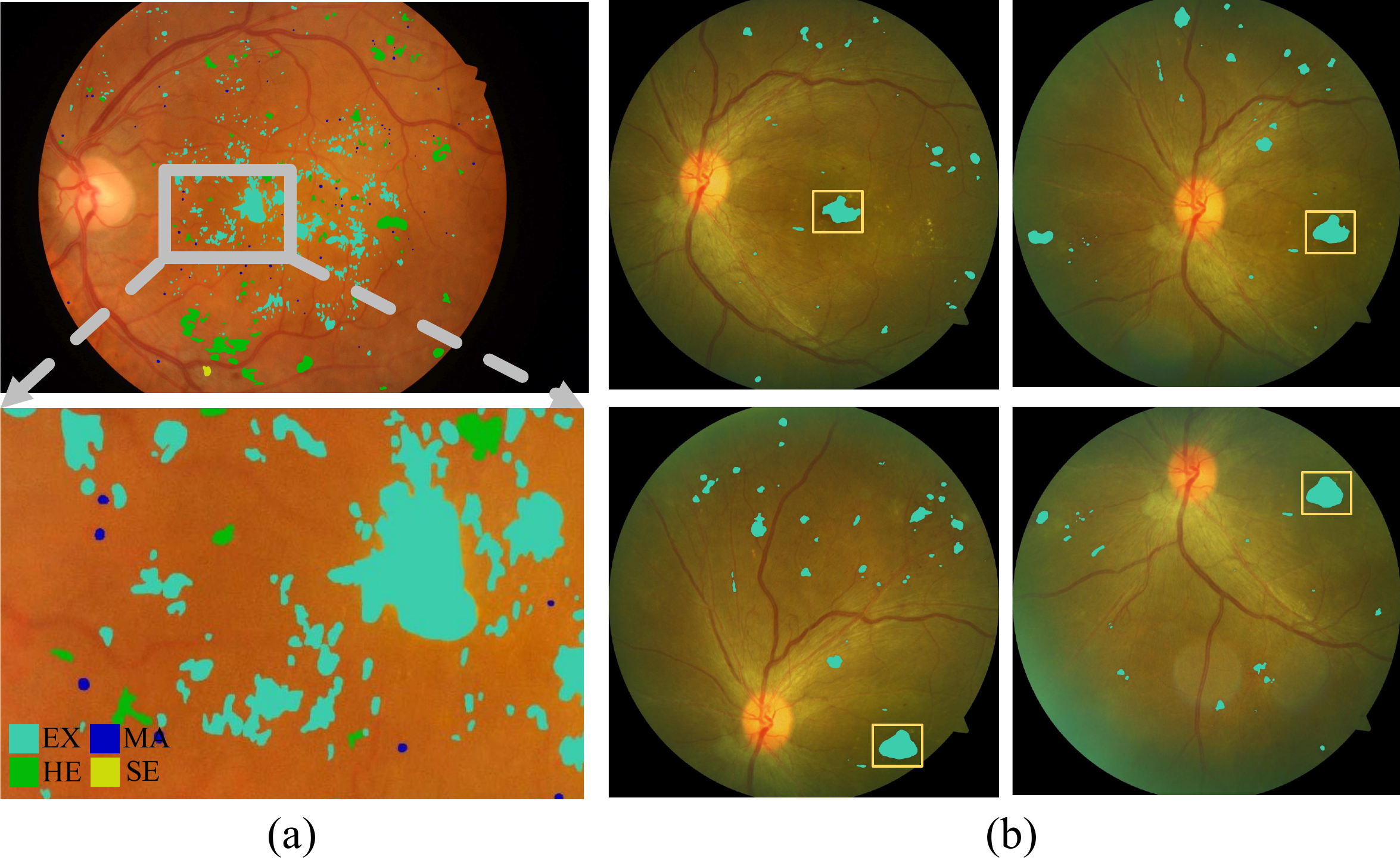}  % 插入图片，设置宽度为页面宽度的50%
	\caption{Samples of color fundus images. (a) Fundus image and zoomed-in region. (b) Multi-view fundus images with lesion distribution. The yellow box encloses the lesions of the same region.}  % 图片标题
	\label{motivation}  % 给图片打上标签，便于引用
	\vspace{-25pt}
\end{figure}
%Samples of fundus images. 

With the advancement of deep learning and its widespread use in various areas\cite{AMSTE,liu2022localized,liu2025reliable}, more researchers are using this technology for automatic DR detection. These methods typically design different network architectures to extract features of fundus images and lesions for DR diagnosis. However, lesions are often vary significantly in size and scattered across the retina, which presents challenges for effective feature learning. As shown in Fig. \ref{motivation}(a), lesion sizes can vary greatly within a single fundus image, making feature extraction particularly difficult. Furthermore, certain types of lesions, such as hemorrhages, tend to be widely distributed with irregular patterns.  For  lesions of different scales, it is crucial for the model to focus on local features to accurately capture subtle changes in the fundus images. For scattered lesions, the model must integrate global contextual information to understand the correlations between lesions across different regions and images. The diversity in lesion size and distribution patterns places demands the model to effectively learn local features and global context for accurate DR detection.

In addition, most existing studies rely on single-view fundus images, covering only 45 to 50 degrees of the retinal field of view, representing about 13\% of the entire retina \cite{luo2023mvcinn}, which can miss critical lesion information. To address this, some researchers have proposed multi-view DR detection methods, using fundus images from different views to obtain more comprehensive pathological information. However, these approaches usually simply splice multi-view features without considering the correlation and redundancy of lesion information between views. As shown in Fig. \ref{motivation}(b), lesions are distributed across all views, indicating long-range dependencies and correlations within the fundus images. Ignoring the correlation between different views will make the feature learning of lesions incomplete. Additionally, the fundus images captured from the four views contain overlapping regions, leading to duplication and redundancy of pathological information. This redundant information will overwhelm the model, making it difficult for the model to distinguish important features. One of the overlapping regions is highlighted in the yellow box in Fig. \ref{motivation}(b). Therefore, simply stitching features from the four views ignores the correlations and redundancies of multi-view lesions, potentially degrading DR detection performance.

In recent years, wavelet transform has gained widespread use in various fields  due to its ability to intuitively capture the texture information at different levels of an image \cite{icme02}. By applying wavelet transform, it is possible to extract critical information from an image, including its overall structure, horizontal and vertical edge details, and high-frequency components at finer levels. In fundus image analysis, the primary challenge in lesion learning lies in accurately identifying the lesion boundaries, which correspond to the high-frequency components of the image. Using wavelet transform to process the lesion features can obtain its high-frequency components, thus effectively identifying its boundary information. However, small lesions often exhibit frequency components that are similar to noise, making it difficult to distinguish them using local feature learning alone. Unlike noise, lesions typically exhibit long-range dependencies within the fundus images. By using these long-range dependencies to enhance the local features, we can significantly improve the learning of small lesions and further distinguish the lesions from noise, which is more conducive to the learning of difficult lesions. 

In this paper, we proposes a novel framework for multi-view DR detection. Specifically, in view of the high demand of  local feature and global context information extraction ability, we designed a double-branch network based on CNN and transformer to learn the global and local features, respectively. Considering the challenges in learning  features of small lesions, we leverage the high-frequency components of the wavelet transform to effectively extract edge features, which are then enhanced by the long-range dependencies from the global branch. In addition, to solve the problem that existing multi-view DR detection methods do not fully integrate features of different views, we designs a Cross-View Fusion Module (CVFM), which learn the correlation of lesions between different views through attention mechanism, and use a learnable query to extract crucial  features from multiple views to effectively reduce the redundancy of features. Our contributions are summarized as follows:

\begin{itemize}
	\item{We proposed a novel method for multi-view DR detection, which use two-branch network to capture local information and global dependence, respectively. Information transmitted from global to local based on wavelet is also used to enhance the learning of different lesions.}
	\item{We introduce a cross-view fusion module, which effectively integrate multi-view information and reduce redundancy through attention mechanism and query learning.}
	\item {Experiments conducted on large datasets demonstrate that the proposed method is both effective and competitive compared to other approaches.}
\end{itemize}

\section{Related work}
\subsection{Single-View Based Methods for DR Detection}
\textbf{CNN-based methods for single view DR Detection.}
% Pratt et al. \cite{PRATT2016200} used the multi-layer CNN network which consists of 10-layer convolutional  and 3-layer fully connected neural network. 
Early researchers usually carry out DR detection based on the existing CNN model. Gulshan et al. \cite {gulshan2016deep} uses the Inception-V3 \cite{Inception} to classify 128,000 retinal images into normal and moderate retinopathy. To remove noise, Mansour et al. \cite{Mansour2018} use Gaussian mixture model and canonical correlation analysis technology to preprocess data and employ AlexNet to extract and classify simple features. Hacisoftaoglu et al. \cite{PRL2019} used the ResNet50\cite{Resnet50} to extractmore complex and rich features, which further improved the detection performance. 
% AlexNet\cite{alexNet}

%With the outstanding performance of transformer in many fields, the DR grading method based on transformer has also been proposed. 
% Huang et al. \cite{huang2024ssit} enhanced the transformer framework by removing irrelevant blocks from the input sequence using saliency mapping, thereby focusing the momentum encoder on salient regions. 
\textbf{Transformer-based methods for single view DR Detection.}
Wu et al. \cite{wu2021vision} divide fundus image into non-overlapping pieces and send them to the transformer model for DR classification. Yao et al. \cite{Yao2022FunSwin} proposed the FunSwin system, which segments the fundus image into equal windows and applies attention within these windows to capture both global and local information. Yang et al. \cite{yang2024novel} applied multi-instance learning to retinal images, using the transformer for feature extraction and a global instance calculation block to aggregate features across instances for better global information extraction.

\vspace{-5pt}
\subsection{Multi-View Methods for Medical Image}
Takahashi et al. \cite {takahashi2017applying} combined color fundus images from four 45° fields across 2740 cases and trained GoogLeNet to classify SDR, PPDR, and PDR. Luo et al. \cite{luo2021pr} collected a four-view dataset and designed the MVDRNet model for multi-view DR grading based on this dataset. Building on this, Luo et al. \cite{luo2023mvcinn} introduced the MVICINN model, which utilized a CNN-Transformer dual-branch network for global and local feature learning, along with a CISAM module for global-local information interaction. To further enhance lesion detection, Luo et al. \cite{LFMVDR} combined fundus images with lesion snapshots and introduced heterogeneous convolution blocks (HCB) and extensible self-attention classes (SSAC).
However, existing approaches often simplify multi-view feature fusion, neglecting the complex correlations between views. In addition, the local and global information processing often emphasizes self-attention within individual branches, the interaction between branches is typically under explored.

%\vspace{-5pt}
\section{Methods}
\begin{figure*}[htbp]  % figure 环境用于插入图片 htbphere, top, bottom, page
	\centering  % 图片居中显示
	\includegraphics[width=0.9\textwidth]{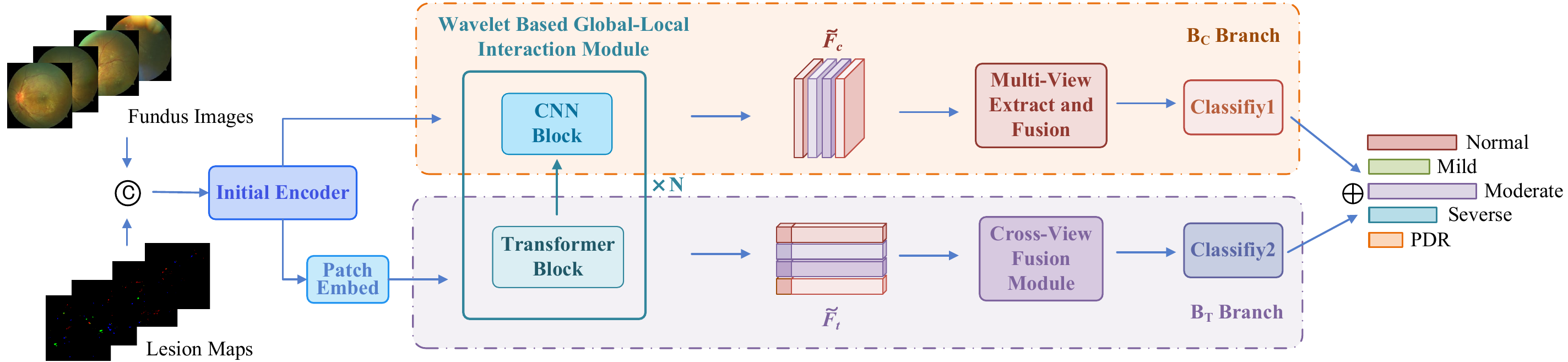}  % 插入图片，设置宽度为页面宽度的50%
	\caption{Overview of the proposed method.}  % 图片标题
	\label{fig_overview}  % 给图片打上标签，便于引用
	\vspace{-15pt}
\end{figure*}
%Our proposed model aims to mine both local and global information from fundus images, and realize the interaction and fusion of them. Therefore, we design a two-branch network to learn the long-distance dependence and local details respectively, and achieve the interaction from global to local through the Wavelet Based Global-Local Interaction Module (WGLIM). Additionally, a Cross-View Attention Module (CVAM) based on attention mechanism and learnable query is introducted to solve the problem of lesion correlation and redundancy between views. This section first introduces the basic knowledge of wavelet transform, followed by the overview of the network's structure. It then provides a detailed description of the proposed WGLIM and CVAM. Finally, we present the classifier and the loss function.

\subsection{Preliminaries: The Wavelet Transform}
Given a image $X\in\mathbb{R}^{C\times H\times W} $ , wavelet transform decomposes the image into four frequency bands: 
\vspace{-5pt}
\begin{equation}
	X_{LL},X_{LH},X_{HL},X_{HH}=DWT(X),
\end{equation}
where $DWT(\cdot)$ denotes the 2D discrete wavelet transform operation. $X_{LL}$, $X_{LH}$, $X_{HL}$ and $X_{HH}$, each of size $\mathbb{R}^{{C}\times {\frac{H}{2}}\times {\frac{W}{2}}}$, capture the following information: $X_{LL}$ captures the overall structure and smooth features of the image. $X_{LH}$, $X_{HL}$ and $X_{HH}$ capture the detail information in the horizontal direction, vertical direction, and the diagonal direction, respectively.

The decomposed subbands can be reconstructed into the original images by inverse wavelet transform  $IWT(\cdot)$  without loss of information, as shown below:
\vspace{-5pt}
\begin{equation}
	X=IWT(X_{LL},X_{LH},X_{HL},X_{HH}).
\end{equation}
\vspace{-20pt}
\subsection{Overview}
The network takes fundus images $X=[X_1,...,X_V]$ and corresponding lesion maps $L=[L_1,...,L_V]$ from multiple views as input, where $V$ is the number of views. For each view $i$, The lesion maps $L_{i}\in\mathbb{R}^{C_l\times H\times W}$ is generated from the fundus image $X_{i}\in\mathbb{R}^{C_x\times H\times W}$ using HACDR-Net\cite{xu2024hacdr}. Both the lesion map $L_{i}$ and the fundus image $X_{i}$ are concatenated together as input $I_{i}$ to the network:
\vspace{-5pt}
\begin{equation}
	I_{i}=Concat(Pre(X_{i}),L_{i}),
\end{equation}
where $I_{i}\in\mathbb{R}^{C\times H\times W}$, and $C=C_x+C_l$. Here, $Pre(\cdot)$ refers to the image preprocessing operation \cite{ratchasima2019} designed to mitigate the effects of lighting conditions. $Concat(\cdot)$ denotes the concatenation operation along the channel dimension.

The multi-view input $I\in\mathbb{R}^{V \times C\times H\times W}$ is first passed through an initial encoder to extract the initial feature $F_c \in\mathbb{R}^{V \times C_c\times H_c\times W_c}$. Considering the similarities across different views, the encoder is shared to reduce the number of model parameters. It consists of a $7\times7$ convolution layer, followed by ReLU activation and max pooling. 

%The use of large convolutional kernels helps capture a broader receptive field, facilitating the extraction of more comprehensive contextual information.

After the initial encoder, the network is divided into two branches: $B_{C}$ (based on CNN) and $B_{T}$  (based on Transformer), which are designed to extract local and global information, respectively. The initial feature map $F_c $ serves as input to the $B_{C}$ branch. Simultaneously, $F_c $ undergoes the PatchEmbed  operation to get the input of $B_T$. The PatchEmbed  first split $F_c $ into multiple patches, which are transformed into the feature $F_{tp} \in\mathbb{R}^{V \times L_p\times D}$, where $L_p$ and $D$ denote the number and dimension of the patches.  $F_{tp}$ are then concatenated with a classification token $F_o\in\mathbb{R}^{V \times {1}\times D}$ to form the input $F_t \in \mathbb{R}^{V \times L \times D}$ for the $B_T$ branch, where $L=L_p+1$.

In both branches, the network first performs common deep feature extraction, followed by separate multi-view feature extraction and fusion. The deep feature extraction stacks $N$ WGLIM, which enable feature interaction between $B_C$ and $B_T$ via wavelet transform. For the $B_C$ branch, the Multi-View Extract and Fusion operation uses convolutional layers to extract features from individual views and then fuse them using a non-local block \cite{wang2018nonlocal}  to model interactions between different views. In the $B_T$ branch, the CVAM is introduced to facilitate multi-view feature interaction and reduce redundancy.

Finally, the features obtained from both branches are passed through a classification header, and the results are combined for the final DR classification.

\subsection{Wavelet Based Global-Local Interaction Module}
\begin{figure}[htbp]  % figure 环境用于插入图片 htbphere, top, bottom, page
	\centering  % 图片居中显示
	\includegraphics[width=0.45\textwidth]{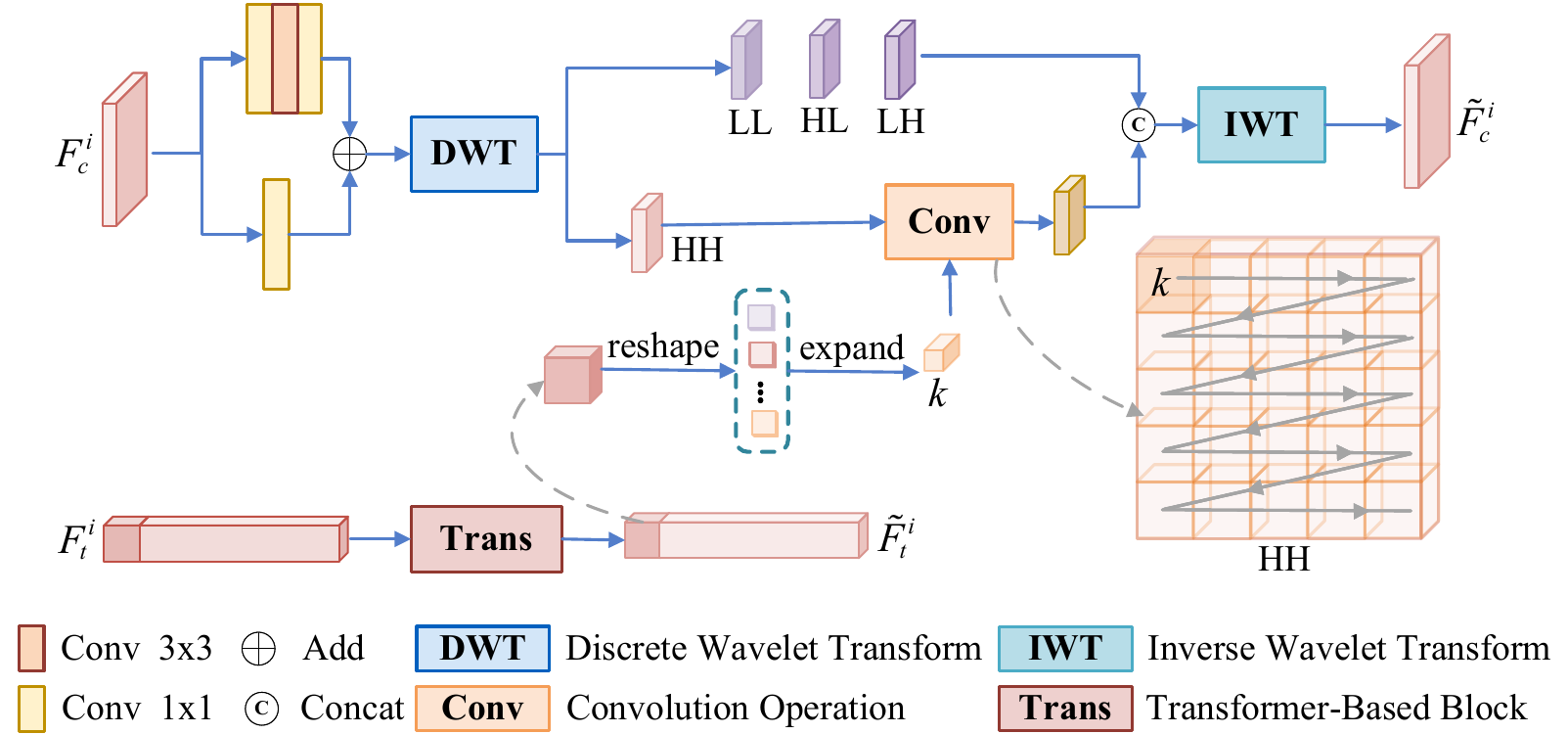}  % 插入图片，设置宽度为页面宽度的50%
	\caption{Structure of the wavelet based global-local interaction module (WGLIM). Take the input of one view as an example.}  % 图片标题
	\label{fig_wmim}  % 给图片打上标签，便于引用
	\vspace{-10pt}
\end{figure}

WGLIM first extracts features and then performs the interaction and fusion of two branches. For feature extraction, the $B_C$ branch learns local information via convolution and employs residual learning \cite{Resnet50} to mitigate network degradation:
\vspace{-5pt}
\begin{equation}
	\hat{F_{c}} = Conv1(F_{c})\oplus Conv2(F_{c}),
\end{equation}
where $\hat{F_{c}} \in\mathbb{R}^{V \times C_c\times H_c\times W_c}$. $Conv1$ represents a convolution operation with three kernels of size $1 \times 1, 3\times 3$ and $1 \times 1$ respectively. $Conv2$ denotes a $1 \times 1$ convolution operation. 

For the $B_T$ branch, a transformer block is used  to extract features from $F_t$. The self-attention mechanism and multi-layer processing of the transformer capture complex dependencies and feature representations. The extracted information is then integrated and compressed into the output features $\tilde{F_{t}}\in\mathbb{R}^{V 
	\times {L}\times D}$. The new classification token $\tilde{F_o}\in\mathbb{R}^{V \times {1}\times D}$ encapsulates the majority of the critical information of $\tilde{F_{t}}$.

%As a digital signal processing method, wavelet is widely used in super-resolution reconstruction \cite{Rehman2018wavelet} and segmentation \cite{finder2024wavelet} due to its strong resolution of signal details and good signal reconstruction quality. Inspired by this, this paper uses wavelet transform to fuse global features and local features. 

Next, global-to-local feature information interaction based on wavelet transform is performed to enhance the learning of local features, particularly those related to lesion edge. Specifically, we apply wavelet transform to the feature $\hat{F_{c}}$ to decompose it into different signal components:
\vspace{-5pt}
\begin{equation}
	F_{c}^{LL},F_{c}^{LH},F_{c}^{HL},F_{c}^{HH}=DWT(\hat{F_{c}}).
\end{equation}

Subsequently, we leverage the global information $\hat{F_o}$ to correct and fuse the local information $F_{c}^{HH}\in \mathbb{R}^{ V \times {C_c }\times \frac{H_c}{2}\times \frac{W_c}{2}}$, which contains the most detailed information among the four components. We use convolution operation for fusion. We first reshape the classification token $\tilde{F_{t}^{o}}\in \mathbb{R}^{V \times 1 \times D}$  into a set of $3 \times 3$ convolution kernels $k_v \in\mathbb{R}^{ V \times C_k\times 3\times 3}$, which is further expanded to $k\in\mathbb{R}^{ V \times {C_c}\times 3\times 3}$ for performing convolutional operations on the high-frequency component $F_{c}^{HH}$:
\begin{equation}
	\hat{F}_{t}^{HH}=Conv(F_{T}^{HH},Expand_{{C_c}}(Reshape_{3\times 3}(\tilde{F_{t}^{o}}))),
\end{equation}
where $Reshape_{3\times 3}{(\cdot)}$ refers to reshaping the feature into multiple $3 \times 3$ convolution kernels. $Conv(a,b)$ represents the convolution operation between $a$ and $b$. $Expand_{C_c}(\cdot)$  denotes the expansion of the feature’s dimension to $C_c$.

Finally, the $\hat{F}_{t}^{HH}$ along with the other components, undergoes an inverse wavelet transform to produce the output $\tilde{F_c}\in\mathbb{R}^{V \times C_c\times H_c\times W_c}$ of $B_C$ branch:
%The wavelet transform and inverse transform enable interaction with global information while preserving the essential local details, thereby improving the extraction of lesion-specific features.
\vspace{-5pt}
\begin{equation}
	\tilde{F_c}=IWT(X_{LL},X_{LH},X_{HL},\hat{F}_{t}^{HH}).
\end{equation}

\subsection{Cross-View Fusion Module}
\begin{figure}[htbp]  % figure 环境用于插入图片 htbphere, top, bottom, page
	\vspace{-5pt}
	\centering  % 图片居中显示
	\includegraphics[width=0.45\textwidth]{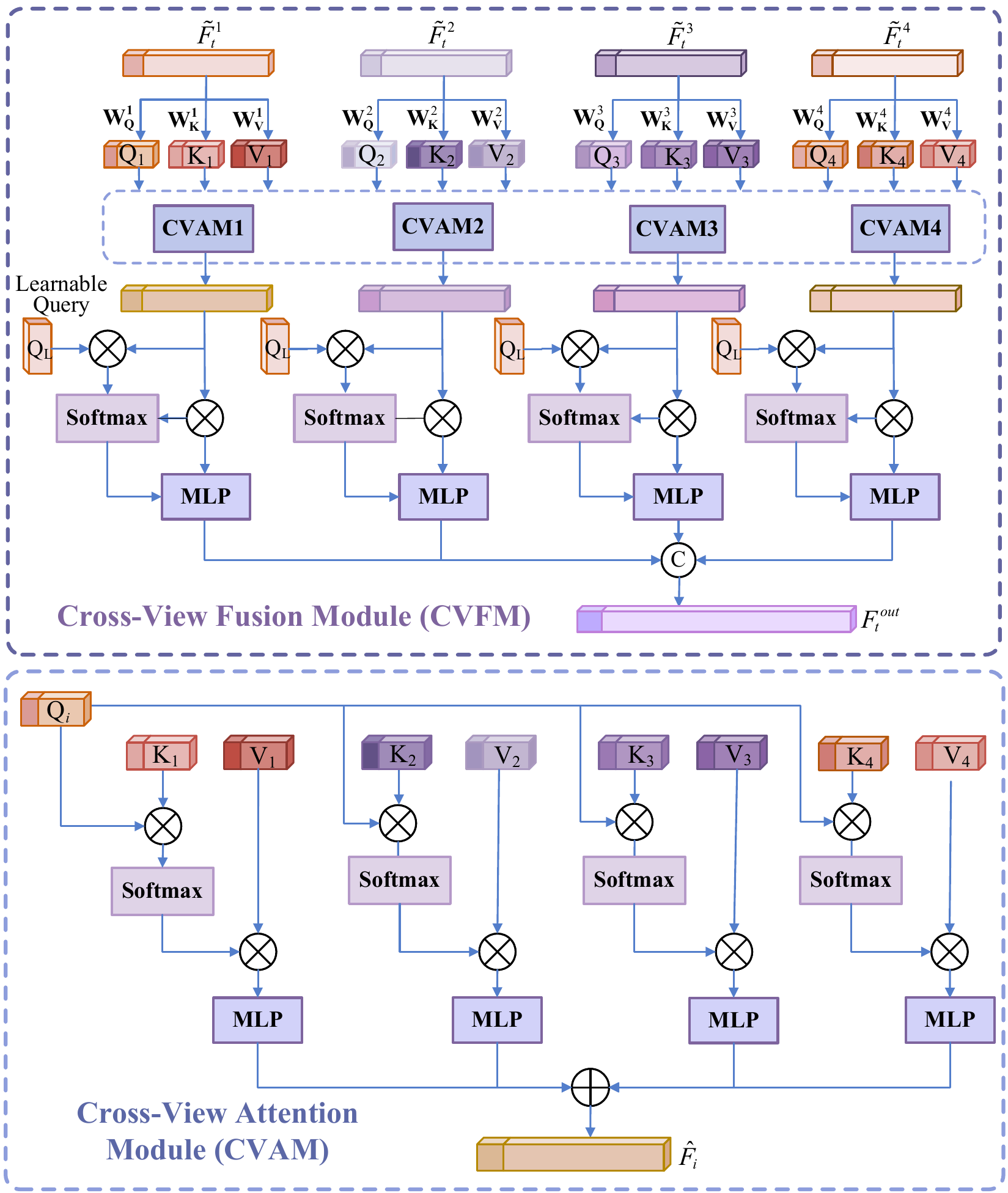}  % 插入图片，设置宽度为页面宽度的50%
	\caption{Structure of th proposed cross-view Fusion module (CVFM).}  % 图片标题
	\label{fig_cvam}  % 给图片打上标签，便于引用
	\vspace{-10pt}
\end{figure}

The CVFM is designed to learn and fuse information from multiple views in two-stage process. 

% The first stage focuses on extracting information from a single view while incorporating complementary information from other views.
\textbf{Stage 1: Learning from a Single View and Incorporating Cross-View Information.}
For each view $i$, the query $Q_i$, key $K_i$, and value $V_i$ are computed from the input $\tilde{F_{t}^i}\in\mathbb{R}^{{L}\times D}$:
\begin{align}
	Q_i =  W_Q^i\tilde{F_{t}^i}, K_i =  W_K^i\tilde{F_{t}^i}, V_i =  W_V^i\tilde{F_{t}^i},
\end{align}
where $W_Q^i\in\mathbb{R}^{D\times D}, W_K^i\in\mathbb{R}^{D\times D},W_V^i\in\mathbb{R}^{D\times D}$ are the learnable weight matrixes. % for the query, key, and value transformations, respectively.

%This approach helps in more effectively capturing cross-view dependencies. 
Next, queries, keys, and values from all views are fed into the Cross-View Attention Module (CVAM). To capture long-range dependencies across views, the model queries not only from the same view but also from other views. The output feature for each view $i$ and view $j$ is computed as:
\vspace{-5pt}
\begin{align}
	atten_{ij} =softmax({Q_iK_j^T}/{\sqrt{d_k}})V_j,    i,j\in[1,2,...,V],  
\end{align}
where $d_k$ is a hyperparameter controlling the scaling factor. 

% , to enhance the model's ability to represent complex information
Later, the features from all views are passed through a multi-layer perceptron  $MLP(\cdot)$, consisting of two linear layers and a ReLU activation function. The outputs are then aggregated across all views:
\vspace{-5pt}
\begin{align}
	\hat{F_i} = {\textstyle \sum_{j=1}^{V}   MLP(atten_{ij}),  i,j\in[1,2,...,V],   } 
\end{align}

At this stage, the outputs from each view contain integrated information from all other views. However, redundancy remains in the fused information.

\textbf{Stage 2: Reducing Redundancy and Refining the Output.} To reduce this redundancy, a learnable query $q\in\mathbb{R}^{L \times D}$ is introduced to attend to the four views, enabling the model to focus on the most relevant information from each view. The query is first applied to each view to compute the attention:
\vspace{-5pt}
\begin{align}
	att_i =softmax({qF_i^T}/{\sqrt{d_k}})\hat{F_i},    i\in[1,2,...,V]. 
\end{align}

The outputs of each view are then passed through an MLP to further refining the representation, which are then concatenated to serves as the final output $F_t^{out} \in\mathbb{R}^{L \times VD}$ of $B_T$ branch :
\vspace{-5pt}
\begin{align}
	F_t^{out} = Concat(MLP(att_1),...,MLP(att_V))).
\end{align}

\vspace{-10pt}
\subsection{Classifier and Loss }

The classifiers in both branches consist of linear layers.  After multi-view feature fusion, the output of \( B_C \) is transformed into \( F_c^{\text{out}} \in \mathbb{R}^{d_c} \), while \( B_T \) uses the output classification token \( F_o^{out} \in \mathbb{R}^{d_t} \) for classification. The final logits \( P_f \in \mathbb{R}^{C} \) are obtained by combining the logits from both branches:
\vspace{-5pt}
\begin{equation}
	P_f = P_c + \alpha P_t,
\end{equation}
where \( P_c \in \mathbb{R}^{C} \) and \( P_t \in \mathbb{R}^{C} \) are the logits of \( B_c \) and \( B_t \). \( \alpha \) is a learnable parameter, and \( C \) is the number of categories.

The loss function is defined as the cross-entropy loss:
\vspace{-5pt}
\begin{align}
	\mathcal{L}= - \textstyle \sum_{f=1}^{C} y_f \log(p_f), \\
	p_f = {\exp(P_f)}/{ \textstyle\sum_{k=1}^{C} \exp(P_k)}, 
\end{align}
where $y_f$ and $p_f$ are the ground truth and predicted probabilities for the $f$-th category.

\section{Experiments}
\subsection{Experimental Setups}

\textbf{Dataset.}
We conducted experiment on MFIDDR \cite{luo2023mvcinn}, a large-scale multi-view fundus image dataset for DR. The dataset collected 34,452 retinal images of 4,334 patients, which were divided into 25,848 training images and 8,604 testing images. Four images were captured at different views for each sample. Seven ophthalmologists rated the DR severity of based on four images, following international standards. All records are anonymous, retrieved with patient consent and hospital approval for data sharing, with no ethical conflicts.

\textbf{Implementation Details.}
The number $N$ of the  WGLIM is set to 8 and the number of views $V$ is 4. The transformer block of the $B_T$ branch is a transformer encoder with 9 heads. We use the Adam optimizer to optimize model, and the initial learning rate is set to $1e-5$. The batch size is set to 24. 

\textbf{Evaluation Metrics.}
We introduce benchmark evaluation metric \cite{trevethan2017sensitivity}, including accuracy (Acc.), precision (Prec.),  specificity (Spec.), Kappa, F1 score (F1) and AUC.

\vspace{-5pt}
\subsection{Comparisons with the Multi-View Methods}

\begin{table}[ht]
	\vspace{-10pt}
	\centering
	\renewcommand{\arraystretch}{1.0}  
	\setlength{\tabcolsep}{3pt}  % 减小列间距
	\caption{Comparison of accuracy(Acc.),  F1 score (F1), kappa and AUC between multi-view methods and our method. The best results are bolded, and the second-best results are underlined. }
	\begin{tabular}{l|ccccc}
		\hline
		Method &Acc. &	Spec.	&Kappa	& {F1}	& AUC \\
		\hline
		Inception\_ResNet\_V2\_MV \cite{Inception}  &	72.57&	85.18	&52.26&	73.22&	89.63\\
		Inception\_V4\_MV  \cite{Inception}&	72.06&	84.27&	51.39	&72.53	&89.04\\
		MobileNet\_V2\_MV \cite{MobileNet}&73.18&	78.25	&49.76	&70.81	&86.04\\
		ResNext50\_32x4d\_MV \cite{Resnet50_32}	&73.50&	\underline{87.20}&	54.85&	74.35	&91.28\\
		VGG16 bn\_MV \cite{vgg16_bn}  &	77.68&	85.48	&59.43	&77.30	&92.15\\
		Swin-B\_MV \cite{swinB}	&78.20&	82.10	&58.94	&77.07	&91.85\\
		Vit-B\_MV \cite{vit}	&78.24	&84.98	&60.07	&77.53	&93.07\\
		MVCINN \cite{luo2023mvcinn}	&80.10	&83.32	&62.45	&78.86	&91.07\\
		LFMVDR-Net \cite{LFMVDR}	&\underline{82.15}	&86.97	&\underline{66.99}	&\underline{81.26}	&\underline{94.81}\\
		Ours & \textbf{84.15}&	\textbf{89.95}&	\textbf{71.16}&	\textbf{83.59} &\textbf{95.14} \\ 
		\hline
	\end{tabular}
	\label{table_all_results}
\end{table}

\begin{table*}
	\vspace{-5pt}
	\setlength{\tabcolsep}{4pt}  % 减小列间距
	\renewcommand{\arraystretch}{1.0}  % 扩大行间距
	\caption{Comparison of precision (Prec.), specificity (Spec.) and  sensitivity and  F-1 score (F1) between multi-view methods and our proposed method. The best results are highlighted in bold, and the second-best results are underlined. (Unit: \%) }
	\begin{tabular}{l|ccc|ccc|ccc|ccc|ccc}
		\hline
		\multirow{2}{*}{ Method}   
		& \multicolumn{3}{c|}{Grade 0}   & \multicolumn{3}{c|}{Grade 1} & \multicolumn{3}{c|}{Grade 2} & \multicolumn{3}{c|}{Grade 3}  & \multicolumn{3}{c}{Grade 4} \\
		\cline{2-16}    
		& Prec. & Spec. & F1 & Prec. & Spec. & F1 & Prec. & Spec. & F1 & Prec. & Spec. & F1 & Prec. & Spec. & F1 \\
		\hline
		Inception\_ResNet\_V2\_MV \cite{Inception} & 88.67 & 82.74 & 85.61 & 48.78 & 82.75 & 54.85 & 52.75 & 95.63 & 52.6 & 64.81 & \textbf{98.1} & 54.69 & 26.19 & 98.53 & 27.16 \\
		Inception\_V4\_MV \cite{Inception} & 87.89 & 81.64 & 84.65 & 46.55 & 81.81 & 52.58 & 51.05 & 95.27 & 52.01 & 66.18 & 97.70 & 63.38 & 66.67 & \underline{99.91} & 17.78 \\
		MobileNet\_V2\_MV \cite{MobileNet} & 82.51 & 68.05 & 87.12 & 59.00 & \textbf{95.19} & 36.48 & 41.20 & 91.51 & 50.11 & 62.20 & 96.9 & 65.38 & 54.55 & 99.76 & 24.00 \\
		ResNext50\_32x4d\_MV \cite{Resnet50_32} & \underline{90.44} & 86.05 & 85.40 & 50.00 & 82.69 & 56.89 & 57.69 & {96.65} & 53.10 & 60.00 & 96.51 & 65.02 & 32.43 & 98.82 & 31.58 \\
		VGG16\_bn\_MV \cite{vgg16_bn} & 88.77 & 81.27 & 89.73 & 58.49 & 88.38 & 60.39 & 53.89 & 95.78 & 53.44 & 66.96 & \textbf{98.1} & 58.56 & 50.00 & 99.62 & 29.09 \\
		Swin-B\_MV \cite{swinB} & 85.92 & 74.91 & 89.68 & 60.89 & 90.73 & 57.81 & 56.46 & \underline{96.75} & 50.30 & \underline{71.43} & \textbf{98.1} & 67.62 & 63.64 & 99.81 & 28.00 \\
		Vit-B\_MV \cite{vit} & 88.34 & 80.17 & 90.12 & 59.33 & 89.26 & 59.53 & 52.87 & 95.83 & 51.54 & 69.92 & \underline{98.00} & 66.19 & \underline{80.00} & \textbf{99.95} & 18.18 \\
		MVCINN \cite{luo2023mvcinn} & 86.71 & \textbf{96.33} & 91.26 & 68.25 & 48.10 & 56.43 & 57.44 & 61.20 & \underline{59.26} & 70.00 & 66.22 & 68.06 & 68.42 & 33.33 & \underline{44.83} \\
		LFMVDR-Net \cite{LFMVDR} & {89.69} & \underline{95.20} & \underline{92.36} & \underline{69.53} & 63.31 & \underline{66.28} & \underline{62.05} & 56.28 & 59.03 & 69.48 & 72.30 & \underline{70.86} & 54.55 & 61.54 & \textbf{57.83} \\
		Ours & \textbf{92.26} & 87.03 & \textbf{93.49} & \textbf{71.02} & \underline{92.31} & \textbf{71.41} & \textbf{63.98} & \textbf{97.05} & \textbf{59.88} & \textbf{71.87} & 97.75 & \textbf{74.68} & \textbf{87.50} & \textbf{99.95} & 29.79 \\
		\hline
	\end{tabular}
	\label{table_complex_results}
	\vspace{-15pt}
\end{table*}
We compare our method with other multi-view approaches, as shown in Tab. \ref{table_all_results}. The compared methods are divided into two categories. The first category includes basic deep learning models, such as Inception\_ResNet\_V2\cite{Inception}, MobileNet\_V2\cite{MobileNet} and Swin-B\cite{swinB}, which take four-view stacks as input. To distinguish them, the suffix "\_MV" is added to each method name. The second category includes existing multi-view methods, such as MVCINN \cite{luo2023mvcinn} and LFMVDR-Net \cite{LFMVDR}. Experimental results show that our method outperforms others in accuracy, specificity, Kappa, F1 score, and AUC, with improvements of 2.10\%, 2.75\%, 4.17\%, 2.33\%, and 0.33\%, respectively, compared to the suboptimal model.

Tab. \ref{table_complex_results} presents results for each DR grade. Our method achieves the highest precision across all grades, demonstrating strong discriminative ability for different DR stages. While some indicators are slightly lower than those of other methods, our approach outperforms or matches the suboptimal performance in 12 of 15 indicators, confirming its superiority over other multi-view methods on the whole.
\vspace{-5pt}
\subsection{Ablation Studies}
%To evaluate the contribution of each module, we designed ablation experiments.  
As shown in Tab. \ref{ablation}, the model proposed is labeled as '$B_c$+$B_t$', where '$B_c$' and '$B_t$' represent the baseline with only the $B_c$ and $B_t$ branch, respectively. The results labeled as  'w/o WGLIM',  'w/o CVAM' refer to methods where WGLIM and CVAM are removed from the proposed model. And '$B_t$ w/o means CVAM is removed from the '$B_t$' baseline. 

\begin{table}[ht]
	\centering
	\setlength{\tabcolsep}{4.5pt}  % 减小列间距
	\renewcommand{\arraystretch}{0.9}  
	\caption{Ablation Studies of our proposed method. The best results are highlighted in bold.}
	\begin{tabular}{l|ccccc}
		\hline
		Method & Acc.(\%) & Prec.(\%) & Spec. (\%)& Kappa(\%) & F1(\%) \\ 
		\hline
		$B_c$ & 75.50 & 78.03 & 88.09 & 57.95 & 75.67 \\
		$B_t$ & 80.38 & 79.49 & 86.58 & 63.97 & 79.20 \\
		$B_c$+$B_t$ & \textbf{84.15} & \textbf{83.95} & \textbf{89.95} & \textbf{71.16} & \textbf{83.59} \\
		$B_t$ w/o CVAM & 64.62 & 44.74 & 51.68 & 17.21 & 52.80 \\
		w/o WGLIM & 82.15 & 82.85 & 89.06 & 67.79 & 81.58 \\
		w/o CVAM & 79.78 & 78.06 & 85.36 & 62.63 & 78.27 \\
		\hline
	\end{tabular}
	\label{ablation}
	\vspace{-10pt}
\end{table}

\textbf{Effectiveness of $B_c$ Branch.} The $B_c$ branch focus on local information extraction. It firstly uses residual convolutional networks to obtain shared features from multiple viewpoints, and then extracts individual viewpoint features and fuses them with non-local block\cite{wang2018nonlocal}. As shown in the Tab. \ref{ablation}, the performance achieved by using this branch alone is unsatisfactory.

\textbf{Effectiveness of $B_t$ Branch.} The branch of $B_t$ focuses on global dependency learning using a transformer block for feature extraction and CAFM for multi-view fusion. As shown in the second row of the Tab. \ref{ablation}, its performance exceeds that of the $B_c$  branch alone, suggesting that long-range dependencies are more crucial for DR detection.

\textbf{Effectiveness of WGLIM.} WGLIM is used to fuse the local and global branches, enhancing detection accuracy. As shown in the Tab. \ref{ablation}, the use of WGLIM leads to improvements in all metrics, demonstrating the effectiveness of the WGLIM.

\textbf{Effectiveness of CVFM.} CVAM is used to perform fusion and reduce redundancy on global features from multiple views. We remove it from both the proposed method and the $B_c$ baseline to evaluate its effectiveness. Adding CVAM improves performance by 15.76\% on the $B_t$ baseline and 4.37\% on the proposed method, highlighting the effectiveness of the CVFM.

\textbf{Effectiveness of Two Stages of CVFM.} We also conducted ablation experiments on the two stages of CVAM, as shown in Fig. \ref{cvam}. The results indicate that using only Stage 1 or Stage 2 leads to performance degradation, suggesting that both stages are essential for improving detection accuracy. These stages work together to enhance multi-view fusion and reduce redundancy through cross-attention  and query learning.

\begin{table}[ht]
	\centering
	\setlength{\tabcolsep}{4.5pt}  % 减小列间距
	\renewcommand{\arraystretch}{1.0}  
	\caption{Comparison of different multi-view fusion methods. The best results are highlighted in bold.}
	\begin{tabular}{l|ccccc}
		\hline
		Methods & Acc.(\%) & Prec.(\%) & Spec. (\%)& Kappa(\%) & F1(\%) \\ 
		\hline
		Stage 1	&83.17&	83.09&	89.34	&69.41	&82.63 \\
		Stage 2&	82.80&	82.88&	\textbf{90.20}&	69.02&	82.64\\
		Stage 1 + Stage 2&\textbf{84.15} & \textbf{83.95} & 89.95 & \textbf{71.16} & \textbf{83.59}\\
		\hline
	\end{tabular}
	\label{cvam}
	\vspace{-20pt}
\end{table}

\subsection{Comparative studies}
\begin{figure}[htbp]  % figure 环境用于插入图片 htbphere, top, bottom, page
		\vspace{-5pt}
	\centering  % 图片居中显示
	\includegraphics[width=0.5\textwidth]{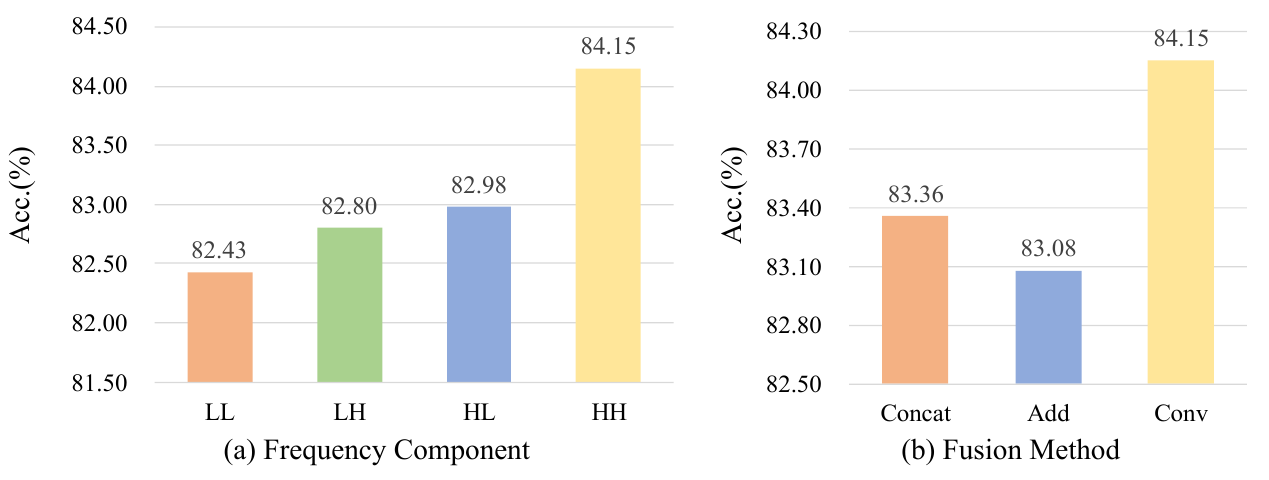}  % 插入图片，设置宽度为页面宽度的50%
	\caption{Comparative studies. }  % 图片标题
	\label{fig_add}  % 给图片打上标签，便于引用
	\vspace{-10pt}
\end{figure}

To further illustrate the effectiveness of using the high-frequency components of the wavelet transform for two-branch information interaction, we replaced the HH components in the model with other components for the study as shown in Tab. \ref{fig_add} (a). Using HH component achieves the highest performance, which shows the its effectiveness.

In addition, we explore different methods for fusing the classification token $\tilde{F_{t}^{o}}$ and high frequency components $F_{c}^{HH}$ as shown in Tab. \ref{fig_add} (b). 'Add' and 'Concat' represent the methods of pixel addition and channel concatenation with $F_{c}^{HH}$ after dimension conversion of $\tilde{F_{t}^{o}}$, respectively. The convolution operation used in this paper achieves the highest performance, illustrating its effectiveness.

\section{Conclusion}
In this paper, we proposed a novel multi-view DR detection framework that addresses challenges in lesion information learning and multi-view fusion. By using a two-branch structure based on CNN and transformer, we capture both local and global features, effectively. We use wavelet transform to enhance the learning of lesion edges and use long-range dependence to improve the feature learning of small lesions. Additionally, the Cross-View Fusion Module (CVFM) is introduced to fuse views and reduce redundancy through attention mechanism and query learning. Experimental results on large datasets demonstrate the effectiveness of our approach.

\bibliographystyle{IEEEtran}
\bibliography{icme2025references}

\end{document}